\documentclass[%
 reprint,
 amsmath,amssymb,
 aps,
 superscriptaddress,
 prd,
]{revtex4-2}

\usepackage[colorlinks,
           bookmarks=true,
           linkcolor=blue,
           urlcolor=blue, 
            anchorcolor=black,
            citecolor=blue
            ]{hyperref}

\usepackage{multirow}
\usepackage{graphicx}
\usepackage{dcolumn}
\usepackage{bm}
\usepackage{float}
\usepackage{exscale}
\usepackage{relsize}

\begin{document}


\title{QCD matter at a finite magnetic field  and nonzero chemical potential}

\author{Zhi-Ying Qin}
\affiliation{School of Physics and Information Technology, Shaanxi Normal University, Xi'an 710119, China}
\affiliation{School of Physics Science and Engineering, Tongji University, Shanghai 200092, China}
\author{Bo Feng}
\affiliation{School of Physics and Information Technology, Shaanxi Normal University, Xi'an 710119, China}
\author{Ya-Hui Hou}
\affiliation{School of Physics and Information Technology, Shaanxi Normal University, Xi'an 710119, China}
\author{Hong-Yue Song}
\affiliation{School of Physics and Information Technology, Shaanxi Normal University, Xi'an 710119, China}
\author{Wen-Chao Zhang}
\email{wenchao.zhang@snnu.edu.cn}
\affiliation{School of Physics and Information Technology, Shaanxi Normal University, Xi'an 710119, China}
\author{Hua Zheng}
\affiliation{School of Physics and Information Technology, Shaanxi Normal University, Xi'an 710119, China}
\author{Shi-Jun Mao}
\affiliation{School of Science, Xi’an Jiaotong University, Xi’an, Shaanxi 710049, China}
\date{\today}

\begin{abstract}

We construct a hybrid equation of state (EoS) by smoothly interpolating the EoS in the hadron resonance gas at low temperatures to that in the ideal parton gas at high temperatures, and employ it to study the properties of the quantum chromodynamics (QCD) matter under finite magnetic field and nonzero chemical potential. In this work, we neglect the anomalous magnetic moment effects of both charged and neutral particles. Our results show that the thermodynamic observables such as the entropy density, the pressure, the energy density, the trace anomaly, and the specific heat  at constant volume are sensitive to both finite magnetic field and chemical potential. As the chemical potential increases from zero, these quantities rise in both the hadronic and quark-gluon plasma phases. In contrast, introducing a magnetic field suppresses them at low temperatures but enhances them at high temperatures.  Furthermore, nonzero chemical potential and magnetic field introduce nontrivial modifications to the squared speed of sound. Both effects increase its value near the critical temperature while reducing it at lower temperatures. When both the chemical potential and the magnetic field are present, their influences superimpose, leading to more intricate changes in the thermodynamic behavior. Finally, we compare our results with the lattice QCD data for the quadratic fluctuations of conserved charges and their correlations. The model successfully reproduces the temperature dependence of these observables at $eB=0$ and 0.04 GeV$^2$. However, at the stronger field strength $eB=0.14$ GeV$^2$, the model underestimates the magnitudes while still capturing the overall temperature trend. 

\end{abstract}

\maketitle

\section{\label{sec:intro}Introduction}

Quantum chromodynamics (QCD) predicts a deconfinement phase transition of the matter from  hadronic phase to  quark-gluon plasma (QGP) phase at
high temperatures or high baryon densities. At low baryon densities and high temperatures, the transition is a smooth crossover \cite{cross-over_transition}. However, at high baryon densities and low temperatures, the transition is of first-order \cite{first_order_phase_transition_2,first_order_phase_transition_3}. QGP is expected to be produced in ultra-relativistic heavy-ion collisions. Exploring the properties of QCD phase transition is one of the main goals in heavy-ion collisions at the BNL Relativistic Heavy Ion Collider (RHIC) and CERN Large
Hadron Collider (LHC).

In non-central heavy-ion collision, an external magnetic field $B$ is expected to be generated by the spectators. It has been shown that at the RHIC energy the magnetic field has the magnitude
of the order of $eB\sim m_{\pi}^2\sim10^{18}$ Gauss, where $e$ is the electric charge, $m_{\pi}$ is the mass of  pions \cite{nocentral_collision_eB_2,nocentral_collision_eB_size_2}. The magnetic field generated  at the LHC energy is around 10 times larger than that at the RHIC energy \cite{nocentral_collision_eB_2,nocentral_collision_eB_size_2}. Such a strong magnetic field can significantly influence the properties of the QCD phase transition and the equation of state (EoS) \cite{QCD_properties_eB_2,QCD_properties_eB_3}. Several models have been applied to investigate these effects. They include the MIT bag model \cite{bag_model_2,  bag_model_4, bag_model_5}, the quark-meson model \cite{QM_1, QM_2, QM_3}, the linear sigma model \cite{linear_sigma}, the functional renormalization group approach \cite{FRG}, the Nambu–Jona-Lasinio (NJL) model \cite{NJL_3, NJL_4, NJL_5, NJL_6, NJL_7}, the lattice QCD (LQCD) \cite{LQCD_2, LQCD_3, LQCD_4, LQCD_5, LQCD_6, LQCD_7}, the hadron resonance gas (HRG) model \cite{HRG_1, HRG_2}, and the perturbative QCD (pQCD) \cite{pQCD_cal_1}.

Recently, the fluctuations and correlations of conserved charges such as the net baryon number $B$, the electric charge $Q$, and the strangeness $S$ in the presence of a finite magnetic field have attracted considerable interest. They are sensitive to the system's degrees of freedom and exhibit distinct behaviors in the hadronic and QGP phases \cite{chi_behaviors_hadron_QGP_1,chi_behaviors_hadron_QGP_2,chi_behaviors_hadron_QGP_3, measure_chi_1}. 
Moreover, the fluctuations are expected to be largely enhanced around the critical end point (CEP) \cite{critical_end_point_1}. Thus they have been measured experimentally to probe the CEP \cite{critical_end_point_3,critical_end_point_4}. In the framework of the NJL model \cite{mag_fluct_corre_NJL_1, mag_fluct_corre_NJL_2, NJL_6, NJL_7}, the LQCD  \cite{mag_fluct_corre_LQCD_1, LQCD_5}, the HRG model \cite{mag_fluct_corre_HRG_0,mag_fluct_corre_HRG_1,mag_fluct_corre_HRG_2, mag_fluct_corre_HRG_3}, and the pQCD \cite{pQCD_cal_2}, it has been shown that the finite magnetic field has a nontrivial effect on these fluctuations and correlations.

Compared with other models, the HRG model is a relatively simple approach. It gives a good description of the LQCD data on the thermodynamic observables as well as the fluctuations of conserved charges at vanishing magnetic field in the low temperature region, both at zero \cite{HRG_model_mu=0_1,HRG_model_mu=0_4, HRG_model_mu=0_5} ($\mu= 0$) and nonzero chemical potentials ($\mu\neq 0$) \cite{HRG_model_mu=/0_1,HRG_model_mu=/0_2,HRG_model_mu=/0_3}. In Ref. \cite{HRG_1}, a modified HRG model was  developed to include the effects of finite magnetic field. This model was utilized to investigate the effect of the magnetic field on the QCD EoS \cite{HRG_1, HRG_2} and the fluctuations of conserved charges at vanishing chemical potentials \cite{mag_fluct_corre_HRG_0,mag_fluct_corre_HRG_1,mag_fluct_corre_HRG_2, mag_fluct_corre_HRG_3}. In the high-temperature limit, the ideal parton gas (IPG) model \cite{parton_gas_model_0} was used to describe the thermodynamic observables and the EoS in the QGP phase \cite{parton_gas_model_1, parton_gas_model_2, parton_gas_model_3}. It was also employed to estimate the fluctuations and correlations of conserved charges in the high temperature region \cite{IPG_fluct_1, IPG_fluct_2}. The estimation well described the results from the LQCD calculation when an ideal and massless quark gas was considered \cite{IPG_fluct_2}. In Ref. \cite{mag_fluct_corre_LQCD_1}, this model was generalized to include the effects of finite magnetic field at zero chemical potential.

In this work, in order to investigate the effects of the finite magnetic field as well as the nonzero chemical potential on the QCD matter, we employ the HRG (IPG) model to describe the EoS in the hadronic (QGP) phase and interpolate these two equations of state with a smooth crossover. This interpolation method has previously been employed, for instance, to study the temperature dependence of thermodynamic properties in the chiral limit with massless pions and light quarks  \cite{parton_gas_model_1}. In our implementation, the hadronic sector includes free hadrons and resonances with masses up to 2.5 GeV/c², while the QGP sector consists of massless gluons and massive $u$, $d$, and $s$ quarks. Moreover, anomalous magnetic moment (AMM) effects of both charged and neutral particles are neglected. We will first investigate the QCD matter at the LHC energy where $\mu\approx0$ GeV and $eB \approx 10~m_{\pi}^2$ to explore the effect of the strong magnetic field. Then we will study the QCD matter at the RHIC energy where $\mu\neq0$ and $eB\approx m_{\pi}^2$ to probe both the effects of the finite magnetic field and the nonzero chemical potential. Finally, we will 
present the second-order fluctuations and correlations of conserved charges as a function of temperature calculated from our model and compare them with the results from LQCD. 

This paper is organized as follows. In sect. \ref{sec:calculation}, we describe the HRG model, the IPG model, and construct a QCD EoS under a finite magnetic field and chemical potential to calculate thermodynamic quantities, as well as the fluctuations and correlations of conserved charges. In sect. \ref{sec:results}, we analyze the effects of  chemical potential and magnetic field on the thermodynamic quantities and compare the fluctuations and correlations of conserved charges obtained from our model with LQCD results. Finally, the conclusion is given in sect. \ref{sec:conclusions}.

\section{\label{sec:calculation}The model}

In this section, we construct a hybrid HRG+IPG equation of state under finite magnetic field and chemical potential. In the absence of a magnetic field, the pressure is isotropic. However, in the presence of a magnetic field, the pressure becomes anisotropic, consisting of a parallel component $P_{\parallel}$ and a perpendicular component $P_{\perp}$. Throughout this work, we adopt the $B$-scheme, where the magnetic field $B$ is treated as a fixed external variable. Under this scheme, the grand-canonical potential is $\Omega(B,T,\mu,V) = -T\ln Z(B,T,\mu,V)$. The parallel pressure is then $P_{\parallel} = -\left(\frac{\partial \Omega}{\partial V}\right)_{T,\mu,B}$, and the total magnetization is $\mathcal{M}_B = -\left(\frac{\partial \Omega}{\partial B}\right)_{T,\mu,V}$. The magnetization density is $m_B = \mathcal{M}_B/V$. The perpendicular pressure follows from $P_{\perp} = P_{\parallel} - m_B B$ \cite{HRG_1,Magnetization}. For the HRG and IPG models, we only give the parallel pressures, whose subscript “$\parallel$” is omitted for brevity. For the hybrid HRG+IPG model, we provide both the parallel and perpendicular components.

\subsection{Hadron resonance gas model}
In the HRG model, with the quantum statistics taken into account, the logarithm of the total  grand canonical partition function of a gas of noninteracting free hadrons and resonances with volume $V$ and temperature $T$ is written as \cite{IPG_fluct_1}
\begin{equation}
\textrm{ln}Z^{\rm H}=\sum_{i\in \textrm{mesons}}\textrm{ln}Z_i^{M}+\sum_{i\in \textrm{baryons}}\textrm{ln}Z_i^{B},
\label{eq-the_giant_canonical_partition_function}
\end{equation}
where $\textrm{ln}Z_i^{M/B}$ refers to the logarithm of the partition function for mesonic ($M$) or baryonic ($B$) particle species $i$ and is given by

\begin{equation}\label{eq:nonextensive_pressure}
\textrm{ln}Z_i^{B/M}= \frac{\pm d_iV}{(2\pi)^3}\int d^3p\ln \left[1\pm{\rm exp}\left(-\frac{\varepsilon_i-\mu_i}{T}\right)\right].
\end{equation}
Here $d_i$ and $\mu_i$ are, respectively, the degeneracy and chemical potential of the particle species $i$,  $\mu_i=B_i\mu_B+Q_i\mu_Q+S_i\mu_S$, where $B_i$, $Q_i$, and $S_i$ are, respectively, the baryon number, the electric charge, and the strangeness of the $i$th particle, $\mu_B$, $\mu_Q$, and $\mu_S$ are the corresponding chemical potentials; the upper sign corresponds to baryons (fermions) and the lower sign to mesons (bosons); $\varepsilon_i=\sqrt{m_i^2+|\boldsymbol{p}|^2}$ is the energy of the particle $i$, with $m_i$ and $\boldsymbol{p}$ being, respectively, its rest mass and  momentum. 

In the presence of a constant uniform magnetic field $eB$ pointing in the positive  $z$ direction,  we consider all particles to be point-like \cite{LQCD_6}. All neutral particles are assumed to be unaffected by the magnetic field, and the AMM effects are neglected. Thus the energy levels of neutral particles are $\varepsilon_i^n=\sqrt{m_i^2+|\boldsymbol{p}|^2}$. However, the charged particles would undergo Landau quantization perpendicular to the magnetic field \cite{QCD_properties_eB_2}. Consequently, their energy levels  become \cite{Landau_quanta_3}
\begin{equation}
\varepsilon_{i}^c=\sqrt{p_z^2+m_i^2+2|q_i|eB\left(l+\frac{1}{2}-s_z\right)},
\label{eq:the_energy_charged_particle}
\end{equation} 
where $q_i$, $p_z$, and $s_z$ are, respectively, the electric charge, the momentum, and the $z$ component of the spin $s_i$ for the particle $i$; the non-negative integer $l$  corresponds to the allowed Landau levels. In Eq. (\ref{eq:the_energy_charged_particle}), a  Landé $g$-factor of 2 is assigned to all charged particles with nonzero spin and the AMM effects are ignored.

Once the partition function is known, the pressure of the hadron resonance gas system is obtained by differentiation:
\begin{equation}
P^{\rm H}=T\frac{\partial\ln Z^{\rm H}}{\partial V}=\sum_{i\in \textrm{neutral}}P_i^{n}+\sum_{i\in \textrm{charged}}P_i^{c},
\label{eq:the_pressure}
\end{equation}
where 
\begin{equation}\label{eq:nonextensive_pressure}
P_i^{n}= \frac{\pm d_iT}{(2\pi)^3}\int d^3p\ln \left[1\pm{\rm exp}\left(-\frac{\varepsilon_i^n-\mu_i}{T}\right)\right],
\end{equation}
\begin{equation}\label{eq:nonextensive_pressure}
\begin{aligned}
P_i^{c}= &\frac{\pm d'_i|q_i|eBT}{2\pi^2}\sum_{s_z=-s_i}^{s_i}\sum_{l=0}^{\infty}\int_0^{\infty} dp_z\\
&\times \ln \left[1\pm{\rm exp}\left(-\frac{\varepsilon_i^c-\mu_i}{T}\right)\right], 
\end{aligned}
\end{equation}
where $d'_i$ is the degeneracy factor other than spin. Here the momentum integrals in the pressure for the charged particles have been modified as \cite{HRG_2, mag_fluct_corre_HRG_0}
\begin{equation}
d_i\int\frac{{d}^3p}{(2\pi)^3}\rightarrow\frac{d'_i|q_i|eB}{2\pi}\sum_{s_z=-s_i}^{s_i}\sum_{l=0}^{\infty}\int_{-\infty}^{\infty}\frac{{ d}p_z}{2\pi}.
\label{eq-the_modification}
\end{equation}
With the utilization of the identity 
\begin{equation}
\textrm{ln}(1+x)=\sum_{k=1}^{\infty}(-1)^{k+1}\frac{x^k}{k},
\label{eq-the_modification}
\end{equation}
where $k$ is the sum index in the Taylor expansion and $|x|<1$, the pressures for the neutral and charged  particles can be, respectively, rewritten as \cite{HRG_2,mag_fluct_corre_LQCD_1} 
\begin{equation}
P_{i}^{n}=\frac{d_iT^2m_i^2}{2\pi^2}\sum_{k=1}^{\infty}(\mp1)^{k+1}\frac{e^{k\mu_i/T}}{k^2}K_2\left(\frac{km_i}{T}\right),
\label{eq:the_ideal_gas_limit_1}
\end{equation}
and
\begin{equation}
\begin{aligned}
P_{i}^{c}=&\frac{d'_i|q_i|eBT}{2\pi^2}\sum_{s_z=-s_i}^{s_i}\sum_{l=0}^{\infty}\varepsilon_i^0\sum_{k=1}^{\infty}(\mp 1)^{k+1}\\
&\times\frac{e^{k\mu_i/T}}{k}K_1\left(\frac{k\varepsilon_i^0}{T}\right),
\end{aligned}
\label{eq:the_pressure_charged_particle}
\end{equation}
with $K_1$ and $K_2$ being, respectively, the first-order and second-order modified Bessel functions of the second kind, $\varepsilon_i^0=\sqrt{m_i^2+2|q_i|eB\left(l+1/2-s_z\right)}$. 

According to the thermodynamic relation $dP_i=s_idT+n_id\mu_i+(m_B)_idB$, with $n_i$ and $(m_B)_i$ being, respectively, the particle and magnetization densities, the entropy density of the hadron resonance gas system is 
\begin{equation}
s^{\rm H}=\sum_{i\in \textrm{neutral}}s_i^{n}+\sum_{i\in \textrm{charged}}s_i^{c},
\label{eq:entropy_density}
\end{equation}
where the entropy densities for the $i$-th neutral ($s_i^{n}=\partial P_i^{n}/\partial T$) and charged particles ($s_i^{c}=\partial P_i^{c}/\partial T$) are, respectively, written as 

\begin{equation}
\begin{aligned}
s_{i}^{n}=&\frac{d_im_i^2}{2\pi^2}\sum_{k=1}^{\infty}(\mp1)^{k+1}\frac{e^{k\mu_i/T}}{k^2}\biggl\{2T K_2\left(\frac{km_i}{T}\right)\\
&-k\mu_iK_2\left(\frac{km_i}{T}\right)+\frac{km_i}{2}\biggl[K_1\left(\frac{km_i}{T}\right)\\
&+K_3\left(\frac{km_i}{T}\right)\biggl]\biggl\},
\end{aligned}
\label{eq:the_entropy_density_neutral_particle}
\end{equation}
and
\begin{equation}
\begin{aligned}
s_{i}^{c}=&\frac{d'_i|q_i|eB}{2\pi^2}\sum_{s_z=-s_i}^{s_i}\sum_{l=0}^{\infty}\varepsilon_i^0\sum_{k=1}^{\infty}(\mp1)^{k+1}\frac{e^{k\mu_i/T}}{k}\\
&\times\biggl\{K_1\left(\frac{k\varepsilon_i^0}{T}\right)-\frac{k\mu_i}{T}K_1\left(\frac{k\varepsilon_i^0}{T}\right)\\
&+\frac{k\varepsilon_i^0}{2T}\biggl[ K_0\left(\frac{k\varepsilon_i^0}{T}\right)+K_2\left(\frac{k\varepsilon_i^0}{T}\right)\biggl]\biggl\}, 
\end{aligned}
\label{eq:the_entropy_density_charged_particle}
\end{equation}
with $K_0$ and $K_3$ being the zeroth-order and third-order modified Bessel functions of the second kind, respectively. In the vanishing magnetic field, the entropy density for charged particles is the same as that for neutral particles.

\subsection{Ideal parton gas model}
In the high temperature limit, the grand canonical QCD partition function reduces to that of an ideal gas of quarks and gluons. The dominant excitations in QGP are free massless gluons and massive $u$, $d$ and $s$ quarks and antiquarks. Similar to that in the HRG model, the pressure of the QGP is expressed as $P^{\rm QGP}=P_g+P_q$, where $P_g$ ($P_q$) is the pressure for gluons (quarks and antiquarks). For gluons, the degeneracy is given by $d_g = 2_{\text{spin}} \times (N_c^2 - 1) = 16$, with $N_c = 3$ being the number of colors. For quarks, the degeneracy excluding spin is $d'_q = N_c= 3$. In the vanishing magnetic field, with the application of Eq. (\ref{eq:the_ideal_gas_limit_1}), the pressures for the  massless gluons as well as  massive quarks and antiquarks are, respectively, expressed as \cite{parton_gas_model_0, mag_fluct_corre_LQCD_1}
\begin{equation}
P_g=\frac{8\pi^2}{45}T^4,
\label{eq-the_pressure_gluons}
\end{equation}
and 
\begin{equation}
P_q=\sum_{f=u,d,s,\bar{u},\bar{d},\bar{s}}\sum_{k=1}^{\infty}\frac{3T^2m_f^2}{\pi^2}(-1)^{k+1}\frac{e^{k\mu_f/T}}{k^2}{K}_2\left(\frac{km_f}{T}\right).
\label{eq:the_pressure_quarks}
\end{equation}
Here for a quark of flavor $f$, $m_f$ and $\mu_f$ denote its mass and chemical potential, respectively, with the latter being decomposed as $\mu_f=B_f\mu_B+Q_f\mu_Q+S_f\mu_S$, where $B_f$, $Q_f$, and $S_f$ are  the baryon number, electric charge, and strangeness of the quark, respectively. 

Following the approach used in the HRG model, the entropy density of the QGP system is given by
\begin{equation}
s^{\rm QGP}=s_g+s_q,
\label{eq:entropy_density_QGP}
\end{equation}
where $s_g=\partial P_g/\partial T$ is the gluon entropy density and $s_q=\partial P_q/\partial T$ is the entropy density for quarks and antiquarks. These are expressed as
\begin{equation}
s_g=\frac{32\pi^2}{45}T^3,
\label{eq:entropy_gluon}
\end{equation}
and
\begin{equation}
    \begin{aligned}
    s_q =&\sum_{f=u,d,s,\bar{u},\bar{d},\bar{s}} \frac{3m_f^2}{\pi^2}\sum_{k=1}^{\infty}(-1)^{k+1}\frac{e^{k\mu_f/T}}{k^2}\biggl\{2T K_2\left(\frac{km_f}{T}\right)\\
    &-k\mu_fK_2\left(\frac{km_f}{T}\right)+\frac{km_f}{2}\biggl[K_1\left(\frac{km_f}{T}\right)\\
    &+K_3\left(\frac{km_f}{T}\right)\biggl]\biggl\}.
    \end{aligned}
\label{eq:the_entropy_density_quark_zero_mag}
\end{equation}

In the presence of magnetic field, the pressure for gluons keeps unchanged, while the pressure for quarks and antiquarks becomes 
\begin{equation}
\begin{aligned}
P_q=\sum_{f=u,d,s}\frac{3|q_f|eBT^2}{\pi^2}\left[P_f^{l=0}(B)+P_f^{l\neq0}(B)\right]
\end{aligned}
\label{eq:the_pressure_quarks_eB}
\end{equation}
where $q_f$ is the electric charge of the quark with the flavor $f$, the contributions from  the lowest Landau level (LLL) $P_f^{l=0}$ and the higher Landau levels $P_f^{l\neq0}$ are, respectively, written as
\begin{equation}
P_f^{l=0}(B)=\frac{m_f}{T}\sum_{k=1}^{\infty}\frac{(-1)^{k+1}}{k}\textrm{cosh}\left(\frac{k\mu_f}{T}\right)K_1\left(\frac{km_f}{T}\right),
\label{eq-the_energy_pz_0_quark}
\end{equation}
and 
\begin{equation}
P_f^{l\neq0}(B)=\frac{2}{T}\sum_{l=1}^{\infty}\varepsilon_f^+\sum_{k=1}^{\infty}\frac{(-1)^{k+1}}{k}\textrm{cosh}\left(\frac{k\mu_f}{T}\right)K_1\left(\frac{k\varepsilon_f^+}{T}\right),
\label{eq:the_energy_pz_0_quark}
\end{equation}
with $\varepsilon^+_f=\sqrt{m_f^2+2|q_f|eBl}$. Here, we have ignored the AMM effects of quarks. The derivation of Eq. (\ref{eq:the_pressure_quarks_eB}) is presented in appendix A.

Similarly, in the presence of magnetic field, the entropy density for gluons keeps unchanged, while the entropy density for quarks and antiquarks becomes 
\begin{equation}
s_q =\sum_{f=u,d,s} \frac{3|q_f|eBT^2}{\pi^2}\left[s_f^{l=0}(B) + s_f^{l\neq 0}(B)\right],
\label{eq:entropy_quark}
\end{equation}
where $s_f^{l=0}(B)$ and $s_f^{l\neq0}(B)$ are, respectively, given by 
\begin{equation}
	\begin{aligned}
		s_f^{l=0}(B)=&\frac{m_f}{2T^3}\sum_{k=1}^{\infty}\frac{(-1)^{k+1}}{k}\biggl\{km_f{\rm cosh}\left(\frac{k\mu_f}{T}\right)\\
       &\times \left[K_0\left(\frac{km_f}{T}\right)+K_2\left(\frac{km_f}{T}\right)\right]+2K_1\left(\frac{km_f}{T}\right)\\
	    &\times\left[T{\rm cosh}\left(\frac{k\mu_f}{T}\right)-k\mu_f{\rm sinh}\left(\frac{k\mu_f}{T}\right)\right]\biggl\},
	\end{aligned}
	\label{eq1-10}
\end{equation}
and 
\begin{equation}
	\begin{aligned}
		s_f^{l\neq0}(B)=&\frac{1}{T^3}\sum_{l=1}^{\infty}\varepsilon_f^+\sum_{k=1}^{\infty}\frac{(-1)^{k+1}}{k}\biggl\{k\varepsilon_f^+{\rm cosh}\left(\frac{k\mu_f}{T}\right)\\
  &\times\left[K_0\left(\frac{k\varepsilon_f^+}{T}\right)+K_2\left(\frac{k\varepsilon_f^+}{T}\right)\right]+2K_1\left(\frac{k\varepsilon_f^+}{T}\right)\\
		&\times\left[T{\rm cosh}\left(\frac{k\mu_f}{T}\right)-k\mu_f{\rm sinh}\left(\frac{k\mu_f}{T}\right)\right]\biggl\}.
	\end{aligned}
	\label{eq1-11}
\end{equation}

\subsection{Smooth crossover around the critical
temperature}

At the critical temperature $T_c$, the liberation of color degrees of freedom drives a sharp increase in entropy density, signaling the transition from hadronic phase to QGP phase. To ensure continuity between the entropy densities of the two phases, we make a smooth interpolation of $s(T)$ between the hadronic gas at low $T$ and the QGP at high $T$. The simplest possible parametrization of the entropy density $s(T)$ that satisfies the thermodynamic inequality $\partial s(T)/\partial T\geq 0$  and the third-law behavior $s(T)\rightarrow \textrm{constant}$ as $T\rightarrow0$  is \cite{parton_gas_model_1}
\begin{equation}
s(T)=f(T)s^{\rm H}(T)+[1-f(T)]s^{\rm QGP}(T),
\label{eq-the_entropy_density_mixed_phase}
\end{equation}
where $f(T)$ is a weighting function,
\begin{equation}
f(T)=\frac{1}{2}\{1-{\rm tanh}[(T-T_c)/\Gamma]\},
\label{eq-the_fT}
\end{equation}
with $\Gamma$ setting the width of the phase transition region. $f(T)$ is a decreasing function of $T$. It approaches unity at low temperatures and 0 at high temperatures. Thus, for $T$ satisfying $\lvert T-T_c\rvert >  \Gamma$, the system approaches the hadronic phase below $T_c$ and the QGP phase above $T_c$, which is consistent with the crossover nature of the QCD transition. In our model, the smooth interpolation employs a fixed critical temperature $T_c$, even at a finite magnetic field ($eB\leq 0.14$ GeV$^2$) and non‑zero baryon chemical potential ($\mu_B\approx 2\ T_c$). This approximation is well justified. Although the early LQCD results \cite{Lattice_magnetic_field} showed a decrease in $T_c$  of about a few MeV at $eB\sim$ 0.2–0.4 GeV$^2$, the latest LQCD results in Ref. \cite{LQCD_6} show that for $eB\leq 0.3$ GeV$^2$ the change in $T_c$ is less than 1 MeV.  Moreover, for $\mu_B\leq 2\ T_c$, the shift of the critical temperature with the chemical potential is also mild \cite{Tc_chem_dependence}. Therefore, within the parameter ranges studied in this work, keeping $T_c$ constant provides a sound and practical simplification for the interpolation scheme.

The anisotropic pressures $P_{\parallel}(T)$ and $P_{\perp}(T)$, the energy density $\varepsilon(T)$, the specific heat at constant volume $C_V(T)$, as well as the parallel and perpendicular components of the squared speed of sound  $(c_s^2)_{\parallel}(T)$ and $(c_s^2)_{\perp}(T)$ then can be, respectively, obtained using the following expressions,
\begin{equation}
P_{\parallel}(T)=\int_0^T s(t)dt,
\label{eq:pressure_parallel_mixed_phase}
\end{equation}
\begin{equation}
P_{\perp}(T)=\int_0^T s(t)dt-B\int_0^T \frac{\partial s(t)}{\partial B}dt,
\label{eq:pressure_perp_mixed_phase}
\end{equation}
\begin{equation}
\varepsilon(T)=Ts(T)-P_{\parallel}(T)+\sum_i \mu_i\int_0^T \frac{\partial s(t)}{\partial \mu_i} dt,
\label{eq19}
\end{equation}
\begin{equation}
\begin{aligned}
C_V(T)=\frac{\partial \varepsilon(T)}{\partial T}\big |_V=T\frac{\partial s(T)}{\partial T}+\sum_i \mu_i\frac{\partial s(T)}{\partial \mu_i},
\end{aligned}
\label{eq20}
\end{equation}
\begin{equation}
\begin{aligned}
(c_s^2)_{\parallel}(T)=\frac{\partial P_{\parallel}(T)}{\partial\varepsilon(T)}=\frac{s(T)}{C_V(T)},
\end{aligned}
\label{eq:cs2_parallel}
\end{equation}
\begin{equation}
\begin{aligned}
(c_s^2)_{\perp}(T)=\frac{\partial P_{\perp}(T)}{\partial\varepsilon(T)}=\frac{s(T)}{C_V(T)}-\frac{1}{C_V(T)}\frac{B\partial s(T)}{\partial B},
\end{aligned}
\label{eq:cs2_perp}
\end{equation}
where the index $i$ in the summation runs over all the particle species. 

\subsection{Fluctuations and correlations of conserved charges}
The fluctuations and correlations of conserved charges ($B$, $Q$, and $S$) encode fine details of the equation of state, making them sensitive probes for changes in the degrees of freedom and the QCD phase structure \cite{chi_behaviors_hadron_QGP_1,chi_behaviors_hadron_QGP_2,chi_behaviors_hadron_QGP_3, measure_chi_1}. They are characterized by the corresponding susceptibilities, which are defined as the derivative of the dimensionless pressure with respect to reduced chemical potential \cite{measure_chi_1}, 
\begin{equation}
\hat{\chi}_{ijk}^{BQS}=\frac{\partial^{i+j+k}P/T^4}{\partial\hat{\mu}_B^i\hat{\mu}_Q^j\hat{\mu}_S^k},
\label{eq-the_fluctuations_correlations_of_conserved_charges}
\end{equation}
where $\hat{\mu}_X=\mu_X/T$, $X=B,\ Q,\ S$. In the presence of a magnetic field, we use the parallel pressure $P_{\parallel}$  because it directly corresponds to the thermodynamic potential density in the $B$-scheme.  We only focus on the computation of quadratic fluctuations and correlations, i.e. $i+j+k=2$. For example, for charged hadrons in the HRG model at zero chemical potential, the quadratic fluctuation and correlation of the conserved charges are, respectively, expressed as \cite{mag_fluct_corre_LQCD_1}
\begin{equation}
\hat{\chi}^X_2=\frac{eB}{2\pi^2T^3}\sum_j|q_j|X_j^2\sum_{s_z=-s_j}^{s_j}\sum_{l=0}^{\infty}f(\varepsilon_j^0),
\label{eq-the_quadratic_fluctuations_of_conserved_charges}
\end{equation}
and 
\begin{equation}
\hat{\chi}^{XY}_{11}=\frac{eB}{2\pi^2T^3}\sum_j|q_j|X_jY_j\sum_{s_z=-s_j}^{s_j}\sum_{l=0}^{\infty}f(\varepsilon_j^0),
\label{eq-the_quadratic_correlations_of_conserved_charges}
\end{equation}
where the index $j$ runs over all charged hadrons, $X_j$ and $\ Y_j$ represent $B$, $Q$, or $S$ of the hadron species $j$, $f(\varepsilon_j^0)=\varepsilon_j^0\sum_{k=1}^{\infty}(\mp1)^{k+1}kK_1(k\varepsilon_j^0/T)$, with $\varepsilon_j^0=\sqrt{m_j^2+2|q_j|eB\left(l+1/2-s_z\right)}$. In the IPG model, the quadratic fluctuations of the conserved charges and their correlations in the high temperature limit can  be expressed as follows
\begin{equation}
	\begin{aligned}
		\hat{\chi}_2^{B}=&\frac{eB}{9\pi^2T^3}\sum_{k=1}^{\infty}k(-1)^{k+1}\biggl\{2m_u K_1\left(\frac{km_u}{T}\right)\\
		&+m_d K_1\left(\frac{km_d}{T}\right)+m_s K_1\left(\frac{km_s}{T}\right)\\
        &+2\sum_{l=1}^{\infty}\biggl[2\varepsilon_u^+ K_1\left(\frac{k\varepsilon_u^+}{T}\right)+\varepsilon_d^+  K_1\left(\frac{k\varepsilon_d^+}{T}\right)\\
        &+\varepsilon_s^+ K_1\left(\frac{k\varepsilon_s^+}{T}\right)\biggl]\biggl\},
	\end{aligned}
	\label{eq:chi_2_B}
\end{equation}
\begin{equation}
	\begin{aligned}
		\hat{\chi}_2^{Q}=&\frac{eB}{9\pi^2T^3}\sum_{k=1}^{\infty}k(-1)^{k+1}\biggl\{8m_u K_1\left(\frac{km_u}{T}\right)\\
        &+m_d K_1\left(\frac{km_d}{T}\right)+m_s K_1\left(\frac{km_s}{T}\right)\\
		&+2\sum_{l=1}^{\infty}\biggl[8\varepsilon_u^+ K_1\left(\frac{k\varepsilon_u^+}{T}\right)+\varepsilon_d^+ K_1\left(\frac{k\varepsilon_d^+}{T}\right)\\
       &+\varepsilon_s^+ K_1\left(\frac{k\varepsilon_s^+}{T}\right)\biggl]\biggl\},
	\end{aligned}
	\label{eq:chi_2_Q}
\end{equation}
\begin{equation}
	\begin{aligned}
		\hat{\chi}_2^S=&\frac{eB}{\pi^2T^3}\sum_{k=1}^{\infty}k(-1)^{k+1}\biggl[m_s K_1\left(\frac{km_s}{T}\right)\\
        &+2\sum_{l=1}^{\infty}\varepsilon_s^+ K_1\left(\frac{k\varepsilon_s^+}{T}\right)\biggl],
	\end{aligned}
	\label{eq:chi_2_S}
\end{equation}
\begin{equation}
	\begin{aligned}
		\hat{\chi}_{11}^{BQ}=&\frac{eB}{9\pi^2T^3}\sum_{k=1}^{\infty}k(-1)^{k+1}\biggl\{4m_u K_1\left(\frac{km_u}{T}\right)\\
        &-m_d  K_1\left(\frac{km_d}{T}\right)-m_s K_1\left(\frac{km_s}{T}\right)\\
		&+2\sum_{l=1}^{\infty}\biggl[4\varepsilon_u^+ K_1\left(\frac{k\varepsilon_u^+}{T}\right)-\varepsilon_d^+ K_1\left(\frac{k\varepsilon_d^+}{T}\right)\\
        &-\varepsilon_s^+ K_1\left(\frac{k\varepsilon_s^+}{T}\right)\biggl]\biggl\},
	\end{aligned}
	\label{eq:chi_2_BQ}
\end{equation}
\begin{equation}
	\begin{aligned}
		\hat{\chi}_{11}^{QS}=&\frac{eB}{3\pi^2T^3}\sum_{k=1}^{\infty}k(-1)^{k+1}\biggl[m_s K_1\left(\frac{km_s}{T}\right)\\
       &+2\sum_{l=1}^{\infty}\varepsilon_s^+ K_1\left(\frac{k\varepsilon_s^+}{T}\right)\biggl],
	\end{aligned}
	\label{eq:chi_2_QS}
\end{equation}
\begin{equation}
	\begin{aligned}
		\hat{\chi}_{11}^{BS}=&-\frac{eB}{3\pi^2T^3}\sum_{k=1}^{\infty}k(-1)^{k+1}\biggl[m_s K_1\left(\frac{km_s}{T}\right)\\
  &+2\sum_{l=1}^{\infty}\varepsilon_s^+ K_1\left(\frac{k\varepsilon_s^+}{T}\right)\biggl],
	\end{aligned}
	\label{eq:chi_2_BS}
\end{equation}
where $m_{u/d/s}$ is the mass of the $u/d/s$ quark, $\varepsilon_u^+=\sqrt{m_u^2+4/3eBl}$, $\varepsilon_{d/s}^+=\sqrt{m_{d/s}^2+2/3eBl}$. The derivations of Eqs. (\ref{eq:chi_2_B}-\ref{eq:chi_2_BS}) are shown in appendix B.

\section{\label{sec:results}Results and discussions}
\subsection{At $eB=10\ m_\pi^2$ and $\mu=0$ MeV }

The QCD matter is initially examined under the conditions of zero chemical potential ($\mu_B=\mu_S=\mu_Q=0$ MeV) and a magnetic field strength of $eB=10\ m_\pi^2$. These values correspond to the estimated chemical potential and magnetic field produced in the off-central Pb-Pb collisions at $\sqrt{s_{\rm NN}}=$ 2.76 and 5.02 TeV \cite{LHC_mag, LHC_chemical}. Within the QGP phase, the $u$, $d$, and $s$ quarks are included, with their masses taken as $m_{u}=m_{d}=3.503\text{ MeV}/c^2$ and $m_{s}=96.4\text{ MeV}/c^2$, respectively \cite{mass_quark}. For the hadronic phase, we consider hadrons and hadronic resonances composed of $u$, $d$, and $s$ quarks that have masses below $2.5$ GeV$/c^2$ \cite{HRG_model_mu=0_4}, as listed by Particle Data Group \cite{PDG2016}.  For the phase transition, $T_c$ is taken to be 156.5 MeV \cite{temperature_T_c}, $ \Gamma$ is set as 0.05 $T_c$, which is identical to the choice made in Ref. \cite{parton_gas_model_1}.

\begin{figure*}[]
\begin{center}
		\includegraphics[scale=1.05]{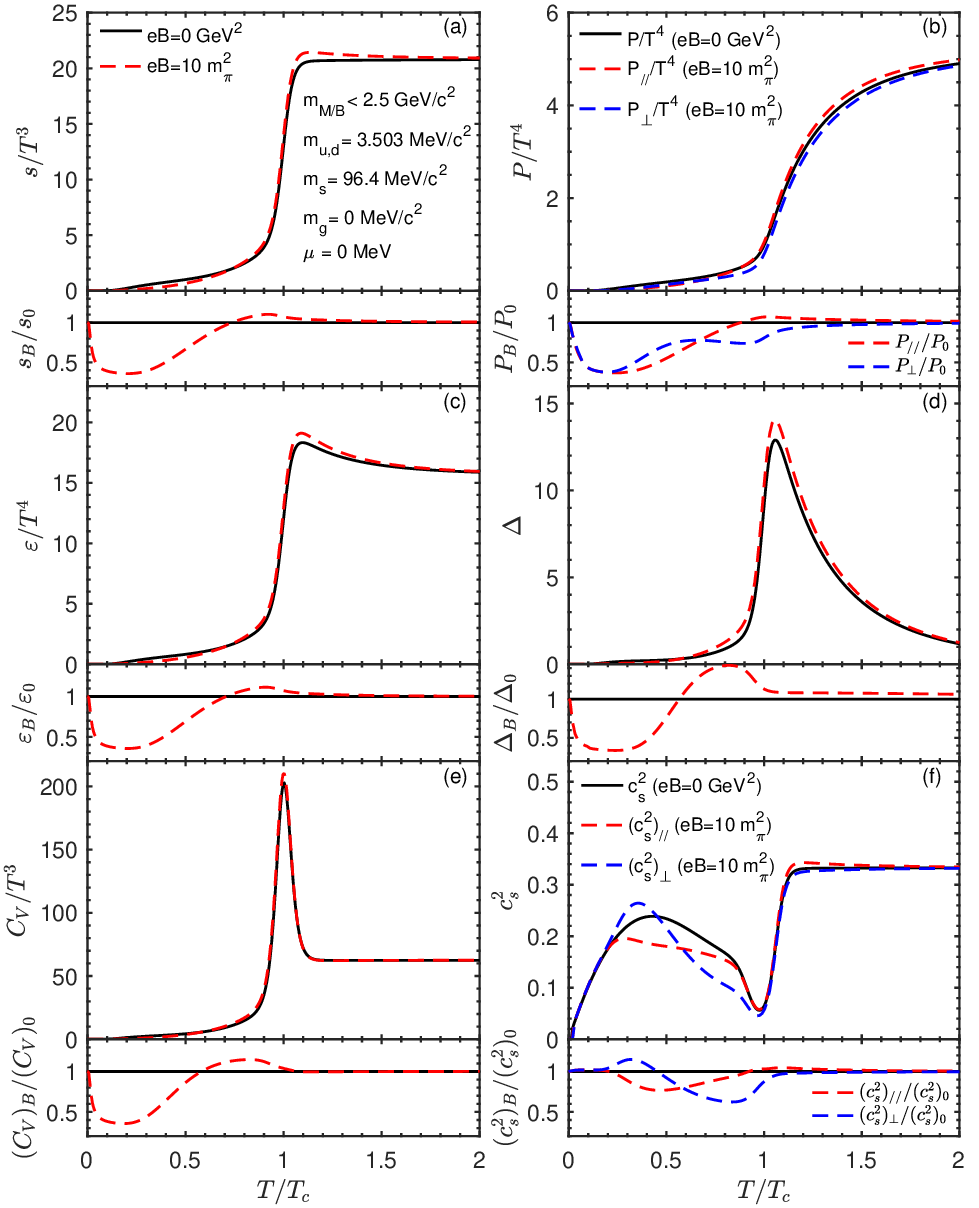}
		\caption{Upper panels in (a) ((b), (c), (d), (e), and (f)): $s/T^3$ ($P_{\parallel}/T^4$, $\varepsilon/T^4$, $\Delta$, $C_V/T^3$, and $(c_s^2)_{\parallel}$) as a function of $T/T_c$.  The dashed red and solid black curves correspond to the cases with $eB =10\ m_{\pi}^2$ and $eB = 0$ GeV$^2$, respectively. In panels (b) and (f), the dashed blue curves correspond to $P_{\perp}/T^4$ and $(c_s^2)_{\perp}$ at $eB =10\ m_{\pi}^2$, respectively. Lower panels: the ratio between the thermodynamic quantities at  $eB =10\ m_{\pi}^2$ and $eB = 0$ GeV$^2$. }
		\label{fig:LHC_eB_new}
\end{center}
\end{figure*}

In the upper panels of each subplot in Fig. \ref{fig:LHC_eB_new}, the red dashed curves present the temperature dependence of the dimensionless thermodynamic quantities, such as the entropy density $s/T^3$, the parallel pressure $P_{\parallel}/T^4$, the energy density $\varepsilon/T^4$,  the trace anomaly $\Delta = [\varepsilon - (P_{\parallel}+2P_{\perp})]/T^4$, the specific heat at constant volume $C_V/T^3$, and the parallel component of the  squared speed of sound $(c_s^2)_\parallel$ at $eB =10\ m_{\pi}^2$. In addition, the blue dashed curves in Fig. \ref{fig:LHC_eB_new} (b) and (f), respectively, show the temperature dependence of  $P_{\perp}/T^4$ and $(c_s^2)_\perp$.  To explicitly highlight the influence of the magnetic field, the corresponding results at $eB = 0$ GeV$^2$  are provided for comparison. They are depicted by solid black curves. In the lower panels of each subplot, we present the ratio of the thermodynamic quantities at $eB =10\ m_{\pi}^2$ to those at $eB = 0$ GeV$^2$. The qualitative behavior of the thermodynamic quantities at finite magnetic field is summarized as follows.

(i) The normalized entropy density $s/T^3$  rises sharply within a narrow critical window of width ($\Gamma$) $\sim$ 8 MeV around the transition temperature $T_c$. At very low temperatures, $s/T^3$ tends asymptotically to zero. Above $T_c$, $s/T^3$ displays a distinct peak before eventually saturating at the Stefan–Boltzmann (SB) limit at very high temperatures. A key observation is that the magnetic field suppresses the entropy density in the low‑temperature hadronic phase but enhances it in the high‑temperature QGP phase. The underlying mechanisms are as follows. At low temperatures, Landau‑level quantization reduces the accessible phase space for charged hadrons, and the increased effective mass raises their excitation threshold, leading to an exponential suppression of their thermal abundance. At high temperatures, the magnetic field reorganizes the phase space of charged quarks. In particular, the LLL   contributes a large degeneracy proportional to the field strength, which significantly increases the number of accessible microscopic states in the QGP phase. The non‑monotonic behavior, specifically the peak just above $T_c$, is driven by the combined effects of the strong magnetic enhancement of the partonic phase and the rapid shift in phase weight governed by the connecting function $f(T)$.

The ratio between the entropy densities with and without the magnetic field, $s_B/s_0$, exhibits the  evolution from extreme low to high temperatures. At very low temperatures, the entropy is  dominated by the lightest neutral hadron $\pi^0$. Contributions from charged pions are exponentially suppressed both with and without the magnetic field, driving $s_B/s_0$ toward unity. As the temperature rises to approximately 0.2 $T_c$, charged pions begin to contribute significantly to the entropy in the zero-field case, whereas in the magnetic field their contribution remains strongly suppressed. This causes the zero-field entropy to accelerate its growth relative to the field-dependent entropy, driving $s_B/s_0$ down to its minimum value. When the temperature increases from 0.2 $T_c$ to around 0.75 $T_c$ where $s_B/s_0=1$, the spectrum of thermally excited hadrons expands to include heavier resonances. The magnetic field suppresses hadrons of different masses to varying degrees: strongly suppressing light hadrons while having a relatively weaker effect on heavier ones. Consequently, its overall suppressive effect gradually diminishes, allowing $s_B/s_0$ to recover and reach unity at 0.75 $T_c$. From 0.75 $T_c$ to 0.9 $T_c$, the ratio $s_B/s_0$ increases continuously from 1 to a peak value of about 1.1. This rise is driven by the rapidly growing influence of the partonic phase, whose entropy is strongly enhanced by the magnetic field. This enhancement overcomes the persistent but diminishing suppression from the hadronic phase, producing a net positive effect that peaks around 0.9 $T_c$. Finally, as the temperature increases from 0.9 $T_c$ to about 2 $T_c$, the ratio $s_B/s_0$ declines from its peak and asymptotically approaches 1. This high‑temperature behavior reflects the transition of the magnetic‑field effect from a non‑perturbative reorganization of phase space to a perturbative correction on the ideal gas state. These corrections scale as powers of $(|q|eB/T^2)^n$ with $n\geq 1$ and therefore vanish in the high‑temperature limit \cite{high_temp_limit}.

(ii) At high temperatures, the normalized parallel pressure $P_{\parallel}/T^4$ asymptotically approaches the SB limit, which is consistent with the leading‑order pQCD result \cite{pQCD_pressure}. For $T > T_c$, it exhibits a moderate increase, a trend that qualitatively agrees with pQCD predictions when higher‑order corrections are taken into account. Moreover, the temperature dependence of the ratio between the pressures with and without the magnetic field, $P_{\parallel}/P_0$, is similar to that of $s_B/s_0$. This is a direct manifestation of the fundamental relation $P_{\parallel}(T) = \int_0^T s(t) dt$, which inherently leads to a smoother, more gradual evolution for the pressure \cite{parton_gas_model_1}. This integral relationship further dictates that the temperatures for the minimum and maximum values of  $P_B/P_0$ must be higher than those for $s_B/s_0$. The normalized perpendicular pressure $P_{\perp}/T^4$ exhibits a temperature dependence similar to that of $P_{\parallel}/T^4$. According to Eq.~(\ref{eq:pressure_perp_mixed_phase}), $P_{\perp}$ is naturally smaller than $P_{\parallel}$. However, at low temperatures, $P_{\perp}$ is slightly larger than $P_{\parallel}$. The underlying mechanism involves the thermal contribution to the magnetization density $m_B$. At very low temperatures, no charged hadrons are thermally excited, so $m_B \approx 0\ \text{GeV}^2$. As the temperature rises, the charged pions, which are scalar bosons, become thermally populated and contribute a negative magnetization. Since $P_{\perp} = P_{\parallel} - m_B B$ (with $B>0\ \text{GeV}^2$), a negative $m_B$ leads to $P_{\perp} > P_{\parallel}$. At higher temperatures, the lightest spin‑1 particle, the $\rho^\pm$ meson, becomes thermally active and provides a positive contribution to $m_B$. This can eventually turn $m_B$ positive, causing $P_{\perp} < P_{\parallel}$ at high temperatures. 


(iii) The temperature dependence of the normalized energy density $\varepsilon/T^4$ rises sharply  near $T_c$ and exhibits a distinct peak just above $T_c$ . This peak arises from the rapid increase of the entropy density $s(T)$ coupled with a relatively slow rise of the pressure in that region \cite{parton_gas_model_1}. It is observed that the ratio between the energy densities with and without a magnetic field $\varepsilon_B/\varepsilon_0$ and $s_B/s_0$ exhibit  similar trends as a function of temperature. It occurs because both the entropy and energy densities are governed by the same underlying  mechanisms and are linked through thermodynamic relations.

(iv) $C_V/T^3$ has an obvious peak near $T_c$.  Its origin lies in the fact that $C_V$ is the derivative of the energy density with respect to temperature, and the energy density undergoes its most rapid, non-linear change  at the critical point. Moreover, the temperature dependence of the ratio between the $C_V$ values with and without the magnetic field, $(C_V)_B/(C_V)_0$, is similar to that of $s_B/s_0$. However, the temperatures for the minimum and maximum values of  $(C_V)_B/(C_V)_0$ are lower than those for $s_B/s_0$. This occurs because $C_V$ is proportional to the temperature derivative of the entropy density, making it more sensitive to the rate of change of $s(T)$ rather than its absolute value.

(v) The trace anomaly  $\Delta$ exhibits a prominent peak just above $T_c$, which reflects the rapid rise in entropy density $s(T)$ associated with the liberation of quark and gluon degrees of freedom during deconfinement. This peak  vanishes as the temperature approaches either zero or infinity. It is observed that the temperature dependence of the ratio between the  $\Delta$ values with and without the magnetic field, $\Delta_B/\Delta_0$, is similar to that of $s_B/s_0$.

(vi) $(c_s^2)_{\parallel}$ exhibits a sudden drop in the critical region $|T - T_c| < \Gamma$. This behavior arises from the energy density $\varepsilon(T)$ rising more rapidly than the parallel pressure $P_{\parallel}(T)$, leading to a soft equation of state.
At low temperatures, $(c_s^2)_{\parallel}$ is  smaller than 1/3, as the energy density $\varepsilon$ receives a significant contribution from the rest masses of hadrons, resulting in $\varepsilon>3P_{\parallel}$. The presence of a magnetic field further suppresses $(c_s^2)_{\parallel}$ by increasing the effective mass of charged hadrons to $m_{\rm eff}=\sqrt{m^2+|q|eB}$, which enhances the exponential Boltzmann suppression $\exp(-m_{\rm eff}/T)$. In the limit $T\rightarrow 0$, both $P_{\parallel}$ and $\varepsilon$ vanish, but their derivatives vanish at different rates, yielding the asymptotic scaling $(c_s^2)_{\parallel}\approx T/m_{\rm eff}$, which guarantees $(c_s^2)_{\parallel}\rightarrow 0$. As the temperature rises above $T_c$, $(c_s^2)_{\parallel}$ recovers and exhibits a peak. This occurs due to a competition between two effects in the deconfined phase: the high effective stiffness contributed by quarks in the LLL and the gradual restoration of three-dimensional ideal gas behavior as higher Landau levels become populated. This competition produces a maximum in $(c_s^2)_{\parallel}$  before it eventually approaches 1/3 at very high temperatures, where the magnetic field’s influence becomes perturbative and the system approaches the  SB limit. According to Eq.~(\ref{eq:cs2_perp}), $(c_s^2)_{\perp}$ is typically smaller than $(c_s^2)_{\parallel}$, yet at low temperatures the former becomes slightly larger. This behavior similar to that of the pressure.


\begin{figure*}[]
\begin{center}
		\includegraphics[scale=1.05]{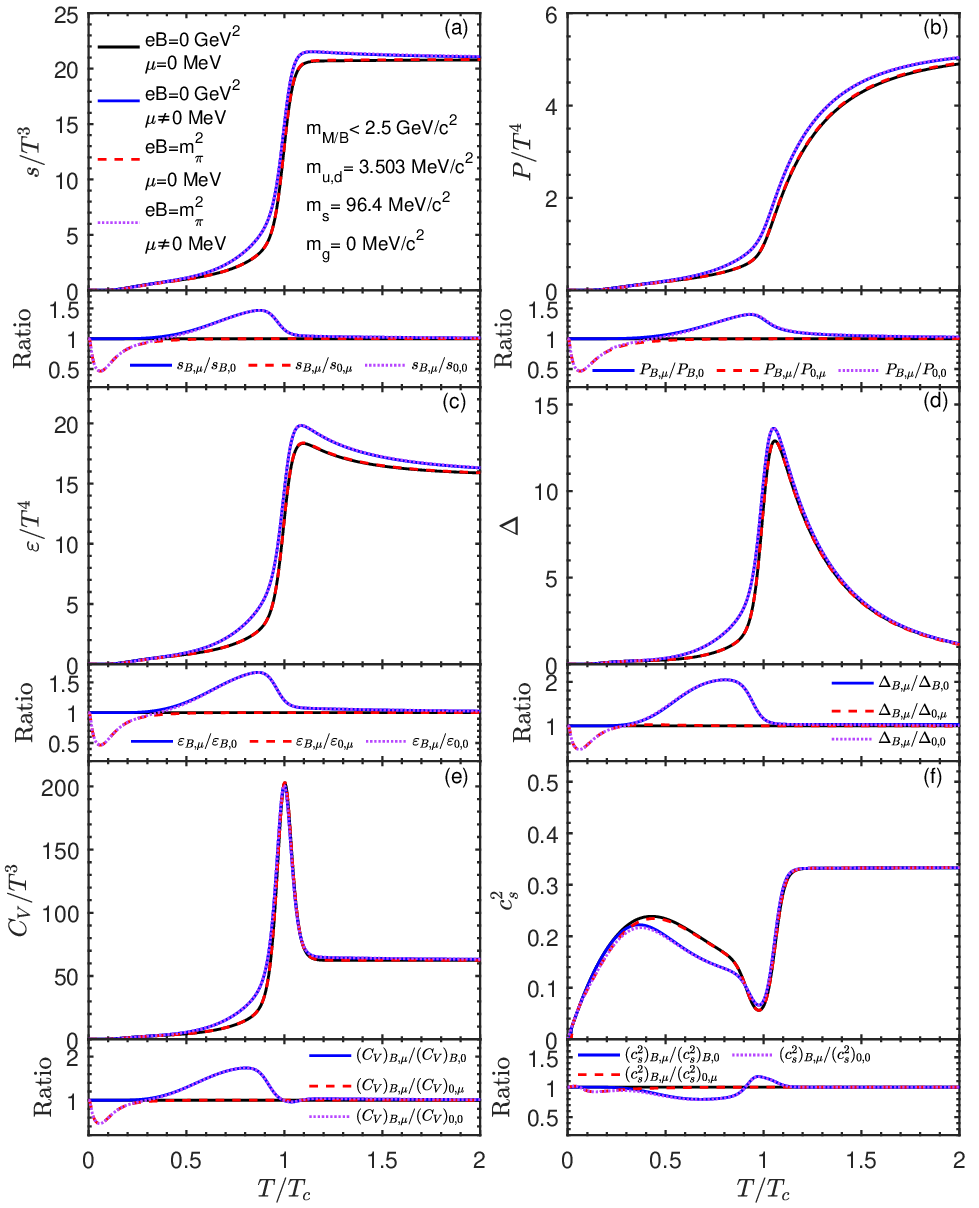}
		\caption{Upper panels in (a) ((b), (c), (d), (e), and (f)): $s/T^3$ ($P/T^4$, $\varepsilon/T^4$, $\Delta$, $C_V/T^3$, and $c_s^2$) as a function of $T/T_c$.  The solid black, solid blue, dashed red, and dotted purple  curves correspond to the cases with $eB = 0$ GeV$^2$ and $\mu=0$ GeV, $eB = 0$ GeV$^2$ and $\mu\neq 0$ GeV,  $eB = m_{\pi}^2$  and $\mu= 0$ GeV, as well as $eB = m_{\pi}^2$  and $\mu\neq 0$ GeV, respectively. Lower panels: the ratios between the thermodynamic quantities denoted by legend.}
		\label{fig:RHIC_eB_new}
\end{center}
\end{figure*}

\subsection{At $eB=\ m_\pi^2$ and $\mu\neq 0$ MeV} 
The QCD matter is further examined under conditions of a finite magnetic field and chemical potential. The field strength is set as $eB=\ m_\pi^2$. The baryon, strangeness, and electric charge chemical potentials are taken as $\mu_B=$ 337.5 MeV, $\mu_S=$ 79.3 MeV, and $\mu_Q=$ 0 MeV, respectively. These values correspond to the estimated magnetic field and chemical potentials produced in off‑central Au–Au collisions at $\sqrt{s_{\rm NN}}=$ 7.7 GeV \cite{RHIC_mag_field, chemical_potential}. Since the magnetic field is weak, the difference between the parallel and perpendicular pressures is small \cite{pressure_anistropy}. Thus we ignore the pressure anisotropy and take the parallel pressure as the pressure of the system. In the upper panels of each subplot in Fig. \ref{fig:RHIC_eB_new}, the solid blue, dashed red, and dotted purple curves present the temperature dependence of the dimensionless thermodynamic quantities with  $eB = 0$ GeV$^2$ and $\mu\neq 0$ MeV,  $eB = m_{\pi}^2$  and $\mu= 0$ MeV, as well as $eB = m_{\pi}^2$  and $\mu\neq 0$ MeV, respectively.  The critical temperature  $T_c$  and $ \Gamma$ are set to the same values used for the case at the LHC energy with $eB = 10\ m_{\pi}^2$ and $\mu= 0$ MeV. For direct comparison, the results for $eB = 0$ GeV$^2$ and $\mu= 0$ MeV are also shown as solid black curves. The lower panels display ratios that highlight the separate and combined effects of the finite magnetic field and chemical potential. The solid blue curves represent the ratio of the thermodynamic quantities at $\mu\neq 0$ MeV to those at $\mu= 0$ MeV, both with $eB = m_{\pi}^2$. The dashed red curves represent the ratios of the thermodynamic quantities at $eB = m_{\pi}^2$ to those at $eB = 0$ GeV$^2$, both with $\mu\neq 0$ MeV. The dotted purple curves represent the ratios of the thermodynamic quantities at $eB = m_{\pi}^2$  and  $\mu\neq 0$ MeV to those at $eB = 0$ GeV$^2$ and  $\mu= 0$ MeV. The qualitative behaviors of the thermodynamic quantities are summarized as follows.

(i) The chemical potential enhances the thermodynamic observables $s/T^3$, $P/T^4$, $\varepsilon/T^4$, $\Delta$, and $C_V/T^3$ in both the hadronic and QGP phases. This enhancement stems from the particle-antiparticle asymmetry induced by the chemical potential. Consequently, the asymmetry increases the accessible microscopic phase space, resulting in an enhancement of these thermodynamic quantities. A consistent trend was observed in the normalized pressure and trace anomaly computed from Dyson-Schwinger equations, where these quantities also grew with increasing quark chemical potential \cite{QCD_PT_chemical_potential}. Moreover, the ratios between these thermodynamic quantities with and without a chemical potential, such as $s_{B,\mu}/s_{B,0}$ and $P_{B,\mu}/P_{B,0}$, display a distinct peak at temperatures below $T_c$. This feature can be understood as a consequence of the chemical potential significantly lowering the effective excitation energy for baryons and strange hadrons, thereby  enhancing their contributions. In contrast, at zero chemical potential, these heavier degrees of freedom remain thermally suppressed. The resulting pronounced enhancement in the presence of $\mu$ leads to a maximum in the ratio. 

(ii) The magnetic field $eB = m_{\pi}^2$ suppresses the thermodynamic observables $s/T^3$, $P/T^4$, $\varepsilon/T^4$, $\Delta$, and $C_V/T^3$ in the low temperature region, which is consistent with the behavior observed for a stronger field $eB=10\ m_\pi^2$.  As the temperature rises to above 0.4-0.5 $T_c$, the thermal energy becomes enough to populate higher Landau levels. Since more levels are occupied, the relative constraining effect of the field on the total phase space weakens. Moreover, more massive neutral resonances are thermally excited, which dilutes the relative impact of the magnetic field on the thermodynamic quantities.  Consequently, the thermodynamic observables calculated with and without the magnetic field begin to converge at this temperature region. Furthermore, the temperature at which the ratio between these thermodynamic quantities with and without the magnetic field reaches a minimum is lower for $eB = m_{\pi}^2$ than that for $eB =10\ m_{\pi}^2$. This shift is due to the competition between thermal excitation and magnetic confinement, governed by the Landau level spacing which scales with $\sqrt{eB}$. For the stronger magnetic field, the larger Landau level spacing requires a higher temperature for thermal energy to effectively overcome the magnetic constraint, shifting the minimum to the higher temperature.

(iii) The squared speed of sound $c_s^2$ exhibits a distinct dependence on the chemical potential. It increases near the critical temperature $T_c$ but decreases at lower temperatures. This behavior can be explained by the way the chemical potential modifies the dominant degrees of freedom in each regime. At low temperatures, a finite chemical potential activates a substantial population of baryons and strange hadrons. These massive degrees of freedom are characterized by a lower $c_s^2$, which reduces the average $c_s^2$ of the system. In contrast, near $T_c$, the chemical potential preferentially enhances the excitation of quark degrees of freedom. This shifts the equation of state toward that of the QGP phase, which possesses a higher characteristic $c_s^2$, thereby increasing $c_s^2$. Additionally, it is observed that the  magnetic field $eB =\ m_{\pi}^2$ suppresses $c_s^2$ at low temperatures, an effect qualitatively similar to that seen under a stronger field $eB =10\ m_{\pi}^2$.

(iv) When both a chemical potential and a magnetic field are present, their effects superimpose, resulting in more complex changes in the thermodynamic behavior. This interplay is reflected in the dotted purple curves shown in the lower panels of each subplot in Fig. \ref{fig:RHIC_eB_new}. These curves display the temperature dependence of the ratios of thermodynamic quantities at $eB = m_{\pi}^2$ and $\mu \neq 0$ MeV to the corresponding quantities at $eB = 0$ GeV$^2$ and $\mu = 0$ MeV.

In Ref. \cite{gamma_dependence}, we have investigated the dependence of thermodynamic quantities on the width parameter $\Gamma$ in the absence of magnetic fields and at zero chemical potential. It was found that increasing $\Gamma$ results in a smoother temperature dependence for both the entropy density and the pressure. Furthermore, the characteristic peaks in the trace anomaly and the specific heat at constant volume become broader and lower in height. For the squared speed of sound, a larger $\Gamma$ leads to a shallower and considerably wider dip. Qualitatively similar trends in the $\Gamma$-dependence are observed even in the presence of a finite magnetic field and non-vanishing chemical potentials.

\begin{figure*}[]
\begin{center}
		\includegraphics[scale=0.98]{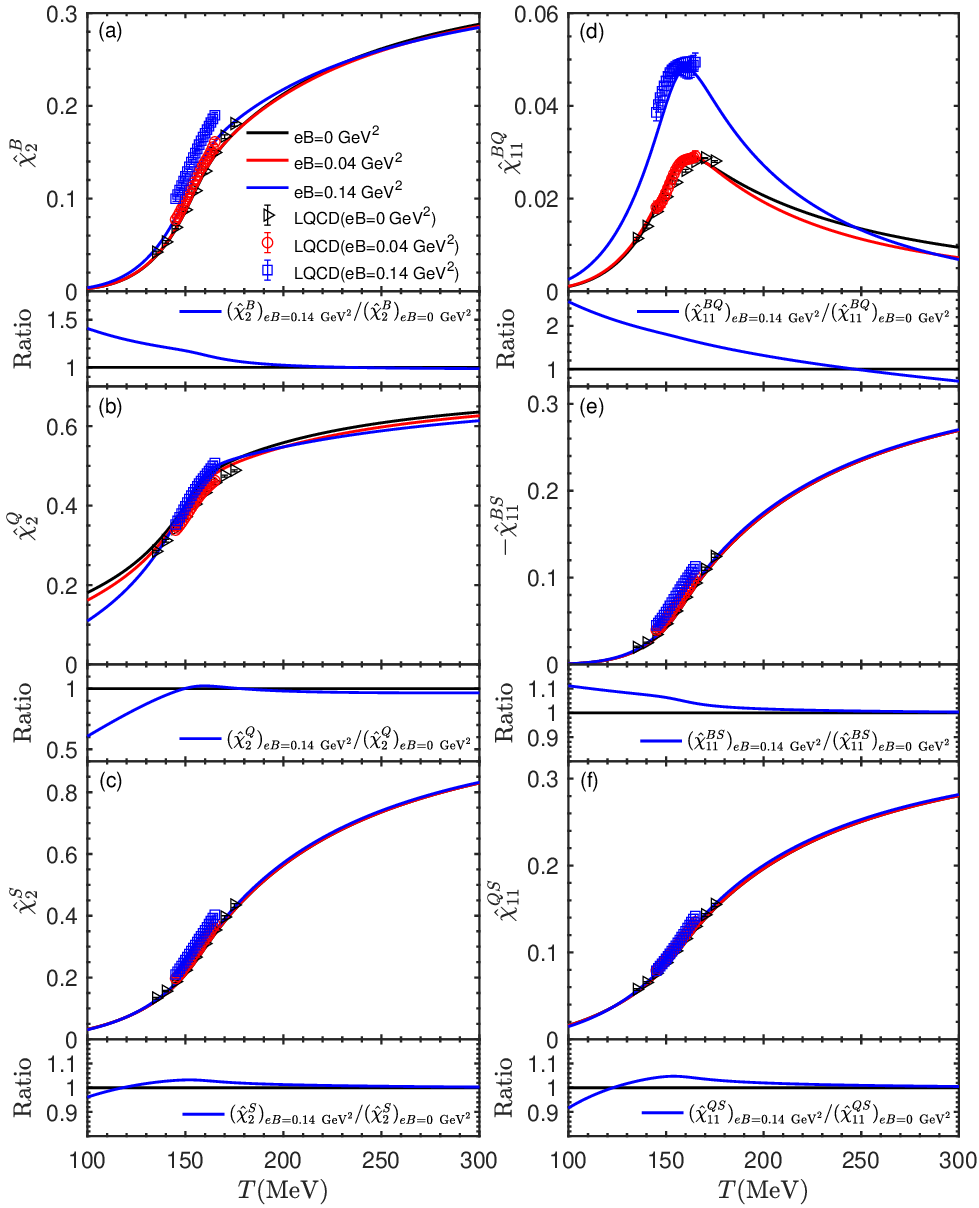}
		\caption{Upper panels in (a) ((b), (c), (d), (e), and (f)): $\hat{\chi}_2^B$ ($\hat{\chi}_2^Q$, $\hat{\chi}_2^S$ , $\hat{\chi}_{11}^{BQ}$, -$\hat{\chi}_{11}^{BS}$, and $\hat{\chi}_{11}^{QS}$) as a function of temperature.  The solid black, red, and blue  curves correspond to the cases with $eB = 0$, 0.04, and 0.14 GeV$^2$, respectively. The LQCD results at these magnetic fields \cite{LQCD_5, LQCD_6} are represented by the empty triangles, circles, and squares, respectively. Lower panels: the ratios
        between the fluctuations or correlations calculated with our model at $eB = 0.14$ and 0 GeV$^2$.}
		\label{fig:luc_corr_BSQ}
\end{center}
\end{figure*}

\subsection{Fluctuations and correlations of conserved charges}
The upper panels of each subplot in Fig. \ref{fig:luc_corr_BSQ} present the temperature dependence of the quadratic fluctuations of conserved charges $B$, $Q$, and $S$, $\hat{\chi}_2^B$, $\hat{\chi}_2^Q$, and $\hat{\chi}_2^S$, as well as  their correlations $\hat{\chi}_{11}^{BQ}$, -$\hat{\chi}_{11}^{BS}$, and $\hat{\chi}_{11}^{QS}$ under different magnetic fields at zero chemical potential. The solid black, red, and blue curves correspond to our model calculations at $eB=$0, 0.04, and 0.14 GeV$^2$,  respectively. For comparison, the LQCD results at the same magnetic fields \cite{LQCD_5, LQCD_6} are displayed using empty triangles, circles, and squares, respectively. The lower panels in each subplot show the ratios between the fluctuations or correlations calculated with our model at $eB = 0.14$ and 0 GeV$^2$. The qualitative behaviors of these fluctuations and correlations can be summarized as follows.

(i) The quadratic fluctuations of conserved charges, $\hat{\chi}_2^B$,  $\hat{\chi}_2^Q$, and  $\hat{\chi}_2^S$, all increase with temperature. This common trend originates from the thermal excitation of charge-carrying degrees of freedom and their dramatic change across the QCD crossover. In the hadronic phase, the dominant carriers differ markedly in mass: the electric charge is carried mainly by the lightest charged pions, the strangeness by heavier kaons, and the baryon number by the heaviest nucleons. Consequently, as temperature rises from zero, $\hat{\chi}_2^Q$ grows first and already reaches a sizable value at moderate temperatures, followed by $\hat{\chi}_2^S$, while $\hat{\chi}_2^B$ remains strongly suppressed and is very small until close to the transition region. This mass‑ordered sequential excitation  is consistent with the HRG picture, in which the equilibrium density of a species is approximately controlled by a Boltzmann factor $e^{-m/T}$. Near  $T_c$, all three susceptibilities exhibit a rapid rise. This steep increase signals the sudden liberation of quark‑level degrees of freedom as the system enters the deconfined regime. In the QGP phase, the effect of quark masses gradually diminishes with increasing temperature, and quarks and antiquarks are more easily excited, causing $\hat{\chi}_2^B$, $\hat{\chi}_2^Q$, and $\hat{\chi}_2^S$ to increase further.  In the high‑temperature limit, they approach the SB values of a non‑interacting three‑flavor quark gas: $\hat{\chi}_2^B$ tends to $1/3$, $\hat{\chi}_2^Q$ tends to 2/3,  and $\hat{\chi}_2^S$ tend to 1.

(ii) The quadratic correlations of conserved charges, $\hat{\chi}_{11}^{QS}$ and $-\hat{\chi}_{11}^{BS}$, increase monotonically with temperature, while $\hat{\chi}_{11}^{BQ}$ shows a non‑monotonic temperature dependence. This pattern can be understood as follows. In the hadronic phase, $\hat{\chi}_{11}^{QS}$ arises mainly from hadrons that carry both $Q$ and $S$, such as  $K^\pm$, $\Sigma^\pm$, and $\Xi^\pm$. Since their $Q$ and $S$ usually have the same sign, they produce a positive correlation. The correlation $\hat{\chi}_{11}^{BS}$ is dominated by strange baryons (e.g., $\Lambda$, $\Sigma$, $\Xi$, $\Omega$), which carry both the  $B$ and  $S$. Their $B$  and $S$ typically have opposite signs, yielding a negative correlation. The main contribution to $\hat{\chi}_{11}^{BQ}$ comes from charged baryons (e.g., $p$, $\Delta^{++}$), for which the $B$ and $Q$ share the same sign, so $\hat{\chi}_{11}^{BQ}>0$. As the temperature rises, more strange and charged hadrons are thermally excited, causing $\hat{\chi}_{11}^{QS}$,  $-\hat{\chi}_{11}^{BS}$ and $\hat{\chi}_{11}^{BQ}$  to grow rapidly. Across the crossover region, quark degrees of freedom begin to be liberated. The strange quark plays a key role because it simultaneously carries electric charge, baryon number, and strangeness. Its contribution is positive for  $\hat{\chi}_{11}^{QS}$ and $\hat{\chi}_{11}^{BQ}$, but negative for $\hat{\chi}_{11}^{BS}$. Compared with hadrons, quarks possess additional color degrees of freedom and have much lighter masses. Consequently, $\hat{\chi}_{11}^{QS}$ and $-\hat{\chi}_{11}^{BS}$ rise sharply during the transition. In the QGP phase, as temperature increases further, the thermal energy  becomes much larger than the strange‑quark mass $m_s$, making strange quarks and antiquarks easier to excite. This enhances $\hat{\chi}_{11}^{QS}$ and $-\hat{\chi}_{11}^{BS}$, which continue to grow with temperature. Meanwhile, the ratio $m_f/T$ decreases for all quark flavors. The quark susceptibilities $\chi_2^u$, $\chi_2^d$, and $\chi_2^s$ all approach unity, but the lighter up and down quarks approach this limit faster than the strange quark. The combination $2\chi_2^u-\chi_2^d-\chi_2^s$, which governs $\hat{\chi}_{11}^{BQ}$, therefore diminishes with rising temperature, causing $\hat{\chi}_{11}^{BQ}$  to decrease after passing through a maximum. In the high‑temperature limit, all three quarks become effectively massless ($m_f/T\rightarrow 0$). This drives $\hat{\chi}_{11}^{QS}$ and $-\hat{\chi}_{11}^{BS}$ to the common SB value of 1/3, while $\hat{\chi}_{11}^{BQ}$ tends to zero.

(iii) The quantities $\hat{\chi}_2^B$, $\hat{\chi}_2^S$, $\hat{\chi}_{11}^{BQ}$, $\hat{\chi}_{11}^{QS}$, and $-\hat{\chi}_{11}^{BS}$ generally increase with magnetic field. However, this trend shows temperature-dependent exceptions: at low temperatures ($T \leq 110\ \text{MeV}\sim 0.7\ T_c$), $\hat{\chi}_2^S$ and $\hat{\chi}_{11}^{QS}$ decrease with the field, while at high temperatures ($T \geq 250\ \text{MeV} \sim 1.6\ T_c$), $\hat{\chi}_2^B$ and $\hat{\chi}_{11}^{BQ}$ also decrease. This behavior can be understood as follows. At low temperatures, the strangeness fluctuations are dominated by the lightest charged kaons. As spin‑0 bosons, kaons experience Landau quantization but have no intrinsic spin magnetic moment. The magnetic field increases the kaon’s effective mass‑squared by $eB$, significantly raising the minimum energy required to produce charged kaons. This suppressive effect overwhelms the increase in Landau‑level degeneracy, leading to a net reduction of $\hat{\chi}_2^S$ and $\hat{\chi}_{11}^{QS}$ with the magnetic field. In the intermediate temperature range, the increase of $\hat{\chi}_2^B$ and $\hat{\chi}_2^S$ is driven by the magnetic field's effect on their primary charged carriers. In addition to the increased low-energy density of states from Landau quantization, charged baryons and many strange hadrons possess spin. For these particles, the coupling of their spin magnetic moment to the field lowers the energy of spin-aligned states. This effectively reduces the energy cost to produce these particles and their antiparticles, thereby enhancing $\hat{\chi}_2^B$ and $\hat{\chi}_2^S$. In the deconfined phase, quarks are similarly affected, contributing to the same trend. The correlations $\hat{\chi}_{11}^{BQ}$, $\hat{\chi}_{11}^{QS}$ and $-\hat{\chi}_{11}^{BS}$ receive dominant contributions from particles that carry both of the relevant charges (e.g., charged baryons for $\hat{\chi}_{11}^{BQ}$, strange mesons for $\hat{\chi}_{11}^{QS}$, strange baryons for $\hat{\chi}_{11}^{BS}$). The magnetic enhancement of these multi‑charge carriers simultaneously boosts their individual abundances and amplifies the correlation between the two conserved charges they carry. Among the correlations, $\hat{\chi}_{11}^{BQ}$ exhibits the largest magnetic field response because the two quantities it correlates, $B$ and $Q$, share charged baryons (e.g., $p$, $\Delta^{++}$, $\Sigma^+$) as their primary carriers. The magnetic field reduces the energy threshold for producing charged baryons and antibaryons, thereby strongly and cooperatively enhancing both baryon-number fluctuations and charge fluctuations, as well as their correlation. At high temperatures, the system undergoes dimensional reduction: the transverse motion of quarks is quantized, and only the LLL remains thermally accessible. Although the LLL possesses a high degeneracy proportional to $eB$, the overall available density of states for charge carriers is drastically reduced compared to the zero‑field, three‑dimensional case. This loss of phase space suppresses the charge fluctuations, causing $\hat{\chi}_2^B$ and $\hat{\chi}_{11}^{BQ}$ to decrease with increasing magnetic field in the high‑temperature regime.

(iv) At low temperatures, $\hat{\chi}_2^Q$ decreases with the magnetic field. Near  $T_c$, it exhibits an increasing trend, while at  high temperatures, it decreases again with the magnetic field. This non-monotonic behavior can be understood as follows. At low temperatures, the charge fluctuations are dominated by the lightest spin-0 charged pions. The reason for the reduction of $\hat{\chi}_2^Q$ with the magnetic field is similar to that of $\hat{\chi}_2^S$. As the temperature approaches $T_c$, heavier charged hadrons with non‑zero spin, such as vector mesons ($\rho^\pm$, $K^{*\pm}$) and baryons($p$, $\Sigma^\pm$), become thermally active and contribute significantly to charge fluctuations. Beyond Landau quantization, they possess a spin magnetic moment. The magnetic field lowers the energy of spin states aligned with it. This enhancing effect from magnetic moment coupling can partially or fully counteract the mass-increase effect from Landau quantization. Consequently, near $T_c$ the magnetic field tends to enhance $\hat{\chi}_2^Q$. At high temperatures, due to the similar reasons as those for $\hat{\chi}_2^B$ and $\hat{\chi}_{11}^{BQ}$,  $\hat{\chi}_2^Q$ decreases with the magnetic field.

(v) Our model successfully reproduces the temperature dependence of the quadratic fluctuations of conserved charges and their correlations calculated by LQCD at $eB=0$ and 0.04 GeV$^2$ \cite{LQCD_5, LQCD_6}. At high temperatures without a magnetic field, the quadratic baryon-number fluctuation from pQCD has been shown to be compatible with LQCD results  \cite{pQCD_fluctuation1}. Our results are therefore consistent with the pQCD predictions in the high-temperature regime. However, at the stronger field strength $eB=0.14$ GeV$^2$, our model underestimates these quantities, despite capturing a similar overall temperature trend. This discrepancy highlights inherent limitations of our model. A key oversimplification is that we assign a  Landé $g$-factor of 2 to all charged particles with nonzero spin and neglect the AMM effects of neutral particles. This is inaccurate for composite hadrons. For instance, the proton has a large anomalous magnetic moment with $g_p \approx 5.586$. In a strong magnetic field, this substantial anomalous moment dramatically enhances the thermal excitation of baryons and antibaryons \cite{anomalous_magnetic_moments}. Our model, with $g=2$, severely underestimates this enhancement, leading to a significant underprediction of baryon-number fluctuations $\hat{\chi}_2^B$ and their correlations with charge, such as $\hat{\chi}_{11}^{BQ}$. Moreover, in the high-temperature region, our model assumes the system is an ideal parton  gas. However, at such strong field strengths, interactions among quarks can become significant. The IPG model  ignores these interaction effects, which are inherently included in LQCD calculations.


\section{\label{sec:conclusions}Conclusions}
Based on a smooth interpolation between the hadron resonance gas and the ideal parton gas, we have developed a hybrid equation of state to investigate the QCD crossover under finite magnetic fields and chemical potentials. Our analysis reveals that the thermodynamic quantities such as the entropy density, the pressure, the energy density, the trace anomaly, and the specific heat  at constant volume  respond sensitively to both external conditions. While a nonzero chemical potential enhances these observables in both phases, a magnetic field suppresses them at low temperatures and enhances them at high temperatures.  Furthermore, both finite density and magnetic field significantly modify the squared speed of sound, increasing it near the transition region but reducing it at lower temperatures. When the chemical potential and magnetic field are present  simultaneously, their effects superimpose, producing a complex thermodynamic response. Moreover, our model successfully reproduces the temperature dependence of conserved charge fluctuations and their correlations from  LQCD at $eB = 0$ and $0.04\ \text{GeV}^2$. However, at the stronger field $eB = 0.14\ \text{GeV}^2$, the model captures the qualitative trend but systematically underestimates the magnitudes. This discrepancy highlights the growing importance of physics beyond our current framework, such as the AMM effects of both charged and neutral particles and the  interactions in the QGP at stronger field strengths. Addressing these effects will be an important direction for our future work.

\begin{acknowledgments}
We would like to thank Jin-Biao Gu and Heng-Tong Ding for providing us the LQCD data. This work is supported by the Scientific Research Foundation for the Returned Overseas Chinese Scholars, State Education Ministry, by Natural Science Basic Research Plan in Shaanxi Province of China (program No. 2023-JC-YB-012), and by the National Natural Science Foundation of China under Grant Nos. 11447024,  11505108 and 12275204.

\end{acknowledgments}
\section*{Appendix A: The derivation of Eq. (\ref{eq:the_pressure_quarks_eB})}\label{Appendix:A}
\setcounter{equation}{0}
\setcounter{subsection}{0}
\renewcommand{\theequation}{A\arabic{equation}}
In the presence of a constant uniform magnetic field $eB$ pointing in the positive  $z$ direction, the pressure for quarks and antiquarks is expressed as
 \begin{equation}
\begin{aligned}
P_q=&\sum_{f=u,d,s,\bar{u},\bar{d},\bar{s}}\frac{3|q_f|eBT}{2\pi^2}\sum_{s_z=-1/2}^{+}\sum_{l=0}^{\infty}\varepsilon_f^0\\
&\times\sum_{k=1}^{\infty}(-1)^{k+1}\frac{e^{\frac{k\mu_f}{T}}}{k}K_1\left(\frac{k\varepsilon_f^0}{T}\right),
\end{aligned}
\label{eq:the_pressure_charged_particle}
\end{equation}
where $\varepsilon^0_f=\sqrt{m_f^2+2|q_f|eB(l+1/2-s_z)}$. For a given flavor $f$, the pressure for the quarks and antiquarks with $s_z=1/2$ is written as
\begin{equation}
\begin{aligned}
P_f^+=&\frac{3|q_f|eBT}{2\pi^2}\biggl\{\sum_{l=0}^{\infty}\varepsilon^{+}_f\sum_{k=1}^{\infty}(-1)^{k+1}\frac{e^{\frac{k\mu_f}{T}}}{k}K_1\left(\frac{k\varepsilon^{+}_f}{T}\right)\\
&+\sum_{l=0}^{\infty}\varepsilon^{+}_f\sum_{k=1}^{\infty}(-1)^{k+1}\frac{e^{-\frac{k\mu_f}{T}}}{k} K_1\left(\frac{k\varepsilon^{+}_f}{T}\right)\biggl\}\\
=&\frac{3|q_f|eBT}{2\pi^2}\sum_{l=0}^{\infty}\varepsilon^{+}_f\sum_{k=1}^{\infty}(-1)^{k+1}\frac{e^{\frac{k\mu_f}{T}}+e^{-\frac{k\mu_f}{T}}}{k}\\ &\times K_1\left(\frac{k\varepsilon^{+}_f}{T}\right)\\
=&\frac{3|q_f|eBT}{\pi^2}\sum_{l=0}^{\infty}\varepsilon^{+}_f\sum_{k=1}^{\infty}\frac{(-1)^{k+1}}{k}{\rm cosh}\left(\frac{k\mu_f}{T}\right)\\
&\times K_1\left(\frac{k\varepsilon^+_f}{T}\right),
\end{aligned}
\label{eq1-5}
\end{equation}
where $\varepsilon^+_f=\sqrt{m_f^2+2|q_f|eBl}$.
For a given flavor $f$, the pressure for the quarks and antiquarks with $s_z=-1/2$ is written as
\begin{equation}
\begin{aligned}
P_f^-=&\frac{3|q_f|eBT}{2\pi^2}\biggl\{\sum_{l=0}^{\infty}\varepsilon^-_f\sum_{k=1}^{\infty}(-1)^{k+1}\frac{e^{\frac{k\mu_f}{T}}}{k} K_1\left(\frac{k\varepsilon^-_f}{T}\right)\\
&+\sum_{l=0}^{\infty}\varepsilon^-_f\sum_{k=1}^{\infty}(-1)^{k+1}\frac{e^{-\frac{k\mu_f}{T}}}{k}K_1\left(\frac{k\varepsilon^-_f}{T}\right)\biggl\}\\
=&\frac{3|q_f|eBT}{2\pi^2}\sum_{l=0}^{\infty}\varepsilon^-_f\sum_{k=1}^{\infty}(-1)^{k+1}\frac{e^{\frac{k\mu_f}{T}}+e^{-\frac{k\mu_f}{T}}}{k} K_1\left(\frac{k\varepsilon^-_f}{T}\right)\\
=&\frac{3|q_f|eBT}{\pi^2}\sum_{l=0}^{\infty}\varepsilon^-_f\sum_{k=1}^{\infty}\frac{(-1)^{k+1}}{k}{\rm cosh}\left(\frac{k\mu_f}{T}\right)K_1\left(\frac{k\varepsilon^-_f}{T}\right)\\
=&\frac{3|q_f|eBT}{\pi^2}\sum_{l=1}^{\infty}\varepsilon^+_f\sum_{k=1}^{\infty}\frac{(-1)^{k+1}}{k}{\rm cosh}\left(\frac{k\mu_f}{T}\right) K_1\left(\frac{k\varepsilon^+_f}{T}\right),
\end{aligned}
\label{eq1-7}
\end{equation}
where $\varepsilon^-_f=\sqrt{m_f^2+2|q_f|eB(l+1)}$.

With the combination of $P_f^+$ and $P_f^-$, the pressure for quarks and antiquarks is rewritten as
\begin{equation}
\begin{aligned}
P_q&=\sum_{f=u,d,s}P_f^+ + P_f^-\\
&=\sum_{f=u,d,s}\frac{3|q_f|eBT^2}{\pi^2}\left[P_f^{l=0}(B)+P_f^{l\neq0}(B)\right],
\end{aligned}
\label{eq:the_pressure_quarks_eB_app}
\end{equation}
where the LLL $P_f^{l=0}$ and the higher Landau levels and $P_f^{l\neq0}$ are, respectively, expressed as
\begin{equation}
P_f^{l=0}(B)=\frac{m_f}{T}\sum_{k=1}^{\infty}\frac{(-1)^{k+1}}{k}\textrm{cosh}\left(\frac{k\mu_f}{T}\right)K_1\left(\frac{km_f}{T}\right),
\label{eq:the_energy_pz_0_quark}
\end{equation}
and 
\begin{equation}
P_f^{l\neq0}(B)=\frac{2}{T}\sum_{l=1}^{\infty}\varepsilon_f^+\sum_{k=1}^{\infty}\frac{(-1)^{k+1}}{k}\textrm{cosh}\left(\frac{k\mu_f}{T}\right)K_1\left(\frac{k\varepsilon_f^+}{T}\right).
\label{eq:the_energy_pz_0_quark}
\end{equation}

\section*{Appendix B: The derivation of Eqs. (\ref{eq:chi_2_B}-\ref{eq:chi_2_BS})}\label{Appendix:B}
\setcounter{equation}{0}
\setcounter{subsection}{0}
\renewcommand{\theequation}{B\arabic{equation}}
By taking the derivatives of the quark pressure in Eq. (\ref{eq:the_pressure_quarks_eB}) with respect to the quark chemical potentials and then setting $\mu_{u, d, s}=0$, one obtains the quark number susceptibilities for the flavors $u$, $d$ and $s$ as follows,
\begin{equation}
	\begin{aligned}
		\chi_2^u=&\frac{\partial^2 P_q}{\partial\mu_u^2}\biggl|_{\mu_{u,d,s}=0}\\
		=&\frac{3\frac{2}{3}eBT}{\pi^2}m_u\sum_{k=1}^{\infty}\frac{(-1)^{k+1}}{k}\frac{k^2}{T^2} K_1\left(\frac{km_u}{T}\right)\\
		&+\frac{6\frac{2}{3}eBT}{\pi^2}\sum_{l=1}^{\infty}\varepsilon_u^+\sum_{k=1}^{\infty}\frac{(-1)^{k+1}}{k}\frac{k^2}{T^2} K_1\left(\frac{k\varepsilon_u^+}{T}\right)\\
		=&\frac{2eB}{\pi^2T}\sum_{k=1}^{\infty}k(-1)^{k+1}\biggl[m_u K_1\left(\frac{km_u}{T}\right)\\
  &+2\sum_{l=1}^{\infty}\varepsilon_u^+ K_1\left(\frac{k\varepsilon_u^+}{T}\right)\biggl],
	\end{aligned}
	\label{eq:u_quark_sus}
\end{equation}
\begin{equation}
	\begin{aligned}
		\chi_2^d=&\frac{\partial^2 P_q}{\partial\mu_d^2}\biggl|_{\mu_{u,d,s}=0}\\
		=&\frac{3\frac{1}{3}eBT}{\pi^2}m_d\sum_{k=1}^{\infty}\frac{(-1)^{k+1}}{k}\frac{k^2}{T^2} K_1\left(\frac{km_d}{T}\right)\\
		&+\frac{6\frac{1}{3}eBT}{\pi^2}\sum_{l=1}^{\infty}\varepsilon_d^+\sum_{k=1}^{\infty}\frac{(-1)^{k+1}}{k}\frac{k^2}{T^2}K_1\left(\frac{k\varepsilon_d^+}{T}\right)\\
		=&\frac{eB}{\pi^2T}\sum_{k=1}^{\infty}k(-1)^{k+1}\biggl[m_d K_1\left(\frac{km_d}{T}\right)\\
  &+2\sum_{l=1}^{\infty}\varepsilon_d^+ K_1\left(\frac{k\varepsilon_d^+}{T}\right)\biggl],
	\end{aligned}
	\label{eq:d_quark_sus}
\end{equation}
\begin{equation}
	\begin{aligned}
		\chi_2^s=&\frac{\partial^2 P_q}{\partial\mu_s^2}\biggl|_{\mu_{u,d,s}=0}\\
		=&\frac{3\frac{1}{3}eBT}{\pi^2}m_s\sum_{k=1}^{\infty}\frac{(-1)^{k+1}}{k}\frac{k^2}{T^2} K_1\left(\frac{km_s}{T}\right)\\
		&+\frac{6\frac{1}{3}eBT}{\pi^2}\sum_{l=1}^{\infty}\varepsilon_s^+\sum_{k=1}^{\infty}\frac{(-1)^{k+1}}{k}\frac{k^2}{T^2} K_1\left(\frac{k\varepsilon_s^+}{T}\right)\\
		=&\frac{eB}{\pi^2T}\sum_{k=1}^{\infty}k(-1)^{k+1}\biggl[m_s K_1\left(\frac{km_s}{T}\right)\\
  &+2\sum_{l=1}^{\infty}\varepsilon_s^+ K_1\left(\frac{k\varepsilon_s^+}{T}\right)\biggl],
	\end{aligned}
	\label{eq:s_quark_sus}
\end{equation}
where $m_{u/d/s}$ is the mass of the $u/d/s$ quark, $\varepsilon_u^+=\sqrt{m_u^2+4/3eBl}$, $\varepsilon_{d/s}^+=\sqrt{m_{d/s}^2+2/3eBl}$.  The mixed quark number susceptibilities $\chi_{11}^{ud}=\chi_{11}^{us}=\chi_{11}^{ds}=0$.

The quark chemical potentials have the following relations with the chemical potentials of baryon number, electric charge and strangeness, 
\begin{equation}
\begin{aligned}
\mu_u =&\frac{1}{3}\mu_B+\frac{2}{3}\mu_Q,\\
\mu_d =&\frac{1}{3}\mu_B-\frac{1}{3}\mu_Q,\\
\mu_s =&\frac{1}{3}\mu_B-\frac{1}{3}\mu_Q-\mu_S.
\end{aligned}
\label{eq:quark_and_conserve_charge_chem_potential}
\end{equation}
Taking the derivative of $P_q$ with respect to the baryon chemical potential and  applying Eq. (\ref{eq:quark_and_conserve_charge_chem_potential}), one obtains
\begin{equation}
\frac{\partial P_q}{\partial \mu_B} = \sum_{i = u,d,s} \frac{\partial P_q}{\partial \mu_i} \cdot \frac{\partial \mu_i}{\partial \mu_B}
      = \frac{1}{3} \left( \frac{\partial P}{\partial \mu_u} + \frac{\partial P}{\partial \mu_d} + \frac{\partial P}{\partial \mu_s} \right).
\end{equation}
Differentiating $P_q$ once more with respect to $\mu_B$, one can derive 
\begin{equation}
\begin{aligned}
\chi_2^B=&\frac{\partial^2 P}{\partial \mu_B^2}\\
= &\frac{1}{3} \bigg[ \left( \frac{\partial^2 P}{\partial \mu_u^2} \frac{\partial \mu_u}{\partial \mu_B} + \frac{\partial^2 P}{\partial \mu_u \partial \mu_d} \frac{\partial \mu_d}{\partial \mu_B} + \frac{\partial^2 P}{\partial \mu_u \partial \mu_s} \frac{\partial \mu_s}{\partial \mu_B} \right) \\
&\quad + \left( \frac{\partial^2 P}{\partial \mu_d \partial \mu_u} \frac{\partial \mu_u}{\partial \mu_B}+ \frac{\partial^2 P}{\partial \mu_d^2} \frac{\partial \mu_d}{\partial \mu_B} + \frac{\partial^2 P}{\partial \mu_d \partial \mu_s} \frac{\partial \mu_s}{\partial \mu_B} \right) \\
&\quad + \left( \frac{\partial^2 P}{\partial \mu_s \partial \mu_u} \frac{\partial \mu_u}{\partial \mu_B} + \frac{\partial^2 P}{\partial \mu_s \partial \mu_d} \frac{\partial \mu_d}{\partial \mu_B} + \frac{\partial^2 P}{\partial \mu_s^2} \frac{\partial \mu_s}{\partial \mu_B} \right) \bigg]\\
=& \frac{1}{3} \cdot \frac{1}{3} \biggl[ \left( \chi^u_2 + \chi_{11}^{ud} + \chi_{11}^{us} \right) 
+ \left( \chi_{11}^{ud} + \chi^d_2 + \chi_{11}^{ds} \right) \\
&+ \left( \chi_{11}^{us} + \chi_{11}^{ds} + \chi^s_2 \right) \biggl] \\
=& \frac{1}{9} \left( \chi^u_2 + \chi^d_2 + \chi^s_2 + 2\chi_{11}^{ud} + 2\chi_{11}^{us} + 2\chi_{11}^{ds} \right)               .
\end{aligned}
\label{eq:chi_2_B_app}
\end{equation}
Similarly, $\chi_2^Q$, $\chi_2^S$, $\chi_{11}^{BQ}$, $\chi_{11}^{QS}$, and $\chi_{11}^{BS}$ are, respectively, expressed as \cite{quark_number_sus}
\begin{equation}
\begin{aligned}
    \chi^Q_2 =& \frac{1}{9} \left( 4\chi^u_2 + \chi^d_2 + \chi^s_2 - 4\chi_{11}^{ud} - 4\chi_{11}^{us} + 2\chi_{11}^{ds} \right), \\
    \chi^S_2 = &\chi_s^2,\\
    \chi_{11}^{BQ} =& \frac{1}{9} \bigl( 2\chi^u_2 - \chi^d_2 - \chi^s_2 + \chi_{11}^{ud} + \chi_{11}^{us} - 2\chi_{11}^{ds} \bigr), \\
    \chi_{11}^{QS} = &\frac{1}{3} (\chi^s_2 - 2 \chi_{11}^{us} + \chi_{11}^{ds}), \\
    \chi_{11}^{BS} =& -\frac{1}{3} \bigl( \chi^s_2 + \chi_{11}^{us} + \chi_{11}^{ds} \bigr).
\end{aligned}
\label{eq:chi_2_Q_S_app}
\end{equation}

Substituting Eqs. (\ref{eq:u_quark_sus})-(\ref{eq:s_quark_sus}) as well as $\chi_{11}^{ud}=\chi_{11}^{us}=\chi_{11}^{ds}=0$ into Eqs. (\ref{eq:chi_2_B_app}) and (\ref{eq:chi_2_Q_S_app}), one gets
\begin{equation}
	\begin{aligned}
		\chi_2^{B}
		=&\frac{eB}{9\pi^2T}\sum_{k=1}^{\infty}k(-1)^{k+1}\biggl\{2m_u K_1\left(\frac{km_u}{T}\right)\\
		&+m_d K_1\left(\frac{km_d}{T}\right)+m_s K_1\left(\frac{km_s}{T}\right)\\
        &+2\sum_{l=1}^{\infty}\biggl[2\varepsilon_u^+ K_1\left(\frac{k\varepsilon_u^+}{T}\right)+\varepsilon_d^+  K_1\left(\frac{k\varepsilon_d^+}{T}\right)\\
        &+\varepsilon_s^+ K_1\left(\frac{k\varepsilon_s^+}{T}\right)\biggl]\biggl\},
	\end{aligned}
	\label{eq:chi_2_B_new_app}
\end{equation}

\begin{equation}
	\begin{aligned}
		\chi_2^{Q}=&\frac{eB}{9\pi^2T}\sum_{k=1}^{\infty}k(-1)^{k+1}\biggl\{8m_u K_1\left(\frac{km_u}{T}\right)\\
        &+m_d K_1\left(\frac{km_d}{T}\right)+m_s K_1\left(\frac{km_s}{T}\right)\\
		&+2\sum_{l=1}^{\infty}\biggl[8\varepsilon_u^+ K_1\left(\frac{k\varepsilon_u^+}{T}\right)+\varepsilon_d^+ K_1\left(\frac{k\varepsilon_d^+}{T}\right)\\
       &+\varepsilon_s^+ K_1\left(\frac{k\varepsilon_s^+}{T}\right)\biggl]\biggl\},
	\end{aligned}
	\label{eq:chi_2_Q}
\end{equation}

\begin{equation}
	\begin{aligned}
		\chi_2^S=&\frac{eB}{\pi^2T}\sum_{k=1}^{\infty}k(-1)^{k+1}\biggl[m_s K_1\left(\frac{km_s}{T}\right)\\
        &+2\sum_{l=1}^{\infty}\varepsilon_s^+ K_1\left(\frac{k\varepsilon_s^+}{T}\right)\biggl],
	\end{aligned}
	\label{eq:chi_2_S_app}
\end{equation}
\begin{equation}
	\begin{aligned}
		\chi_{11}^{BQ}=&\frac{eB}{9\pi^2T}\sum_{k=1}^{\infty}k(-1)^{k+1}\biggl\{4m_u K_1\left(\frac{km_u}{T}\right)\\
        &-m_d  K_1\left(\frac{km_d}{T}\right)-m_s K_1\left(\frac{km_s}{T}\right)\\
		&+2\sum_{l=1}^{\infty}\biggl[4\varepsilon_u^+ K_1\left(\frac{k\varepsilon_u^+}{T}\right)-\varepsilon_d^+ K_1\left(\frac{k\varepsilon_d^+}{T}\right)\\
        &-\varepsilon_s^+ K_1\left(\frac{k\varepsilon_s^+}{T}\right)\biggl]\biggl\},
	\end{aligned}
	\label{eq:chi_2_BQ_app}
\end{equation}
\begin{equation}
	\begin{aligned}
		\chi_{11}^{QS}=&\frac{eB}{3\pi^2T}\sum_{k=1}^{\infty}k(-1)^{k+1}\biggl[m_s K_1\left(\frac{km_s}{T}\right)\\
       &+2\sum_{l=1}^{\infty}\varepsilon_s^+ K_1\left(\frac{k\varepsilon_s^+}{T}\right)\biggl],
	\end{aligned}
	\label{eq:chi_2_QS_app}
\end{equation}
\begin{equation}
	\begin{aligned}
		\chi_{11}^{BS}=&-\frac{eB}{3\pi^2T}\sum_{k=1}^{\infty}k(-1)^{k+1}\biggl[m_s K_1\left(\frac{km_s}{T}\right)\\
  &+2\sum_{l=1}^{\infty}\varepsilon_s^+ K_1\left(\frac{k\varepsilon_s^+}{T}\right)\biggl].
	\end{aligned}
	\label{eq:chi_2_BS_app}
\end{equation}
The dimensionless susceptibilities then can be expressed as
\begin{equation}
\begin{aligned}
    \hat{\chi}^X_2 =& \chi^X_2/T^2,\\
    \hat{\chi}_{11}^{XY} = &\chi_{11}^{XY}/T^2,
\end{aligned}
\label{eq:chi_2_Q_S_app}
\end{equation}
where $X,\ Y= B,\ Q,\ S$.



\end{document}